# CODO: An Ontology for Collection and Analysis of Covid-19 Data

Biswanath Dutta[1][a], Michael DeBellis[b]
[1]*Indian Statistical Institute, Bangalore, India*
[2]*Semantic Web Consultant, San Francisco, CA, USA*
*bisu@drtc.isibang.ac.in, mdebellissf@gmail.com*

Abstract: The COviD-19 Ontology for cases and patient information (CODO) provides a model for the collection and analysis of data about the COVID-19 pandemic. The ontology provides a standards-based open source model that facilitates the integration of data from heterogenous data sources. The ontology was designed by analysing disparate COVID-19 data sources such as datasets, literature, services, etc. The ontology follows the best practices for vocabularies by re-using concepts from other leading vocabularies and by using the W3C standards RDF, OWL, SWRL, and SPARQL. The ontology already has one independent user and has incorporated real world data from the government of India.

[a] https://sites.google.com/site/dutta2005/home
[b] https://www.michaeldebellis.com/

# 1 INTRODUCTION

The COVID-19 pandemic is a worldwide crisis jeopardizing the health of everyone on the planet. One of the tools to combat the pandemic is the collection and analysis of data using FAIR principles.[3] Organizing data with technology based on FAIR principles can provide open, federated data sources that will provide healthcare workers with the critical information required to track and eventually control the growth of the pandemic. The COviD-19 Ontology for cases and patient information (CODO) is a first step at utilizing knowledge graph technology to help combat the pandemic.

There are other initiatives that took a similar approach (discussed in section 2.1). However, CODO is unique in its scope and design approach. The main goals of CODO are to:

1. Serve as an explicit ontology for use by data and service providers to publish COVID-19 data using FAIR principles.
2. Develop and offer distributed, heterogenous, semantic services and applications (e.g., decision support system, advanced analytics).
3. Provide a standards-based reusable vocabulary for the use of various organizations (e.g., government agencies, hospitals, academic researchers, data publishers, news agencies, etc.) to annotate and describe COVID-19 information.

The design of CODO has primarily been motivated by the various COVID-19 data projection websites. For example:

- https://covid19.who.int/
- https://www.isibang.ac.in/~athreya/incovid19/
- https://www.mygov.in/covid-19/
- https://coronavirus.maryland.gov/

These sites show static presentations of COVID-19 cases, patient travel history, the relationships between patients, etc. However, these kinds of static data and visual representations need to be manually processed. The search and visualization capabilities are typically hard coded and impossible for users to customize beyond the parameters defined in the software. More importantly, the data is tightly coupled with specific software to view it.

With the development of the CODO ontology, we aim at supporting the organization and representation of COVID-19 case data on a daily basis, so that the produced data can be queried and retrieved semantically, and can also be taken as an input to carry out advanced analytics (e.g., trend study, growth projection). CODO also aims to facilitate the representation of patient data, the relationships between patients, between patient and locations, changes over time, etc. This network data can support the behaviour analysis of the disease, possible route of disease spreading, various factors of disease transmission, etc.

The CODO ontology will also help policymakers. For example, in analysing how infrastructure was utilized and where infrastructure could have been utilized more effectively. Thus, CODO will help deal with the current pandemic as well as provide a tool to prepare for future potential crises.

The main contributions described in this paper are:

(i) Describe the CODO ontology. How it was developed, how it relates to similar projects, how the ontology can currently be leveraged to support analysis of COVID-19 data and plans for future work.
(ii) Illustrate the process of automatic data integration to the ontology.
(iii) Provide examples of how CODO has already been utilized to analyse data about the pandemic.

The rest of the article is organized as follows: section 2 describes the background that motivated development of CODO. Specifically, a survey of related work, an overview of FAIR principles and how knowledge graphs can be utilized to provide technology that implements these principles. Section 3 describes the methodology used to design the CODO ontology. Section 4 describes the CODO ontology highlighting some of the significant aspects of it. Section 5 evaluates the CODO ontology by automatically loading data on the pandemic and by describing SPARQL queries that can analyse the data. Finally, section 6 concludes the paper and discusses next steps.

[3] https://www.go-fair.org/fair-principles/

# 2 BACKGROUND

In this section we describe related work that we surveyed before developing CODO. We also describe the FAIR principles that were a driving rationale for our decision to use knowledge graph technology.

## 2.1 Related Work

Dealing with a global pandemic is a knowledge intensive process. As a result there have been several ontologies developed related to the COVID-19 pandemic. Before developing CODO we did a survey to determine if we could re-use an existing ontology. We found nine relevant ontologies. However, none of them were in the same space as what we needed: to provide a semantic layer on top of case data from India and the world. We briefly describe some of the other COVID-19 ontologies in this section. Currently, we have not found publications for any of them except for the CIDO ontology (He et al., 2020).

The CIDO ontology (Ontology of Coronavirus Infectious Disease) is part of the OBO Foundry Ontology Library. CIDO is focused on analysing Covid-19 from a medical standpoint. E.g., similarity to other viruses, common symptoms, drugs that have been attempted to treat the virus, etc.

COVID-19 Surveillance Ontology[4] is an application ontology designed to support surveillance in primary care. The main goal of this ontology is to support COVID-19 cases and related respiratory conditions using data from multiple brands of computerized medical record systems. This work is partially related to CODO. However, this ontology is designed as a taxonomy consisting of classes such as education for COVID-19, exposure to COVID-19, definite and possible COVID-19, etc. This ontology does not consist of any properties. This reduces the semantic expressivity of the ontology.

DRUGS4COVID19[5] defines medications and their relationships related to COVID-19. Some of the key classes of the ontology are drug, effect, disease, symptoms, disorder, chemical substance, etc. OVID-19[6] is an ontology that consists of classes to enable the description of COVID-19 datasets in RDF. Some of the classes of this ontology are Dataset, Dataset of the Johns Hopkins University, etc.

The World Health Organization's (WHO) COVIDCRFRAPID[7] ontology is a semantic data model for the WHO's COVID-19 RAPID case record form from 23 March 2020. This model provides semantic references to the questions and answers of the form.

The two ontologies that come closest to CODO are Kg-COVID-19[8] (KG hub to produce a knowledge graph for COVID-19 and SARS-COV-2.) and Linked COVID-19 Data: Ontology[9]. However, both of these ontologies have little semantic information in OWL and are dependent on specific additional software to utilize them.

CODO is an ontology that represents COVID-19 case data in a format based only on OWL and other W3C standards which can be utilized by both other ontologies and software systems. CODO provides tracking of specific cases of the pandemic with details such as how the patient is thought to have been infected and potential additional contacts who may be at risk due to their relationship to the infected individual. CODO also provides tracking of clinical tests, travel history, available resources, and actual need (e.g., ICU bed, invasive ventilators), trend study and growth projections.

## 2.2 FAIR Principles

The FAIR principles (Wilkinson 2016) are widely seen as the best practice for scientific data. These principles require that data be:

- Findable. Data must have rich metadata and unique and persistent identifiers.
- Accessible. Metadata and data should be understandable both to humans and machines.
- Interoperable. Data and metadata should use standards based languages that facilitate the use of automated reasoning and federated queries.
- Reusable. Data should leverage open industry standard technology and domain vocabularies.

## 2.3 Knowledge Graphs

[4] https://bioportal.bioontology.org/ontologies/COVID19
[5] https://github.com/oeg-upm/drugs4covid19-kg
[6] http://covid19.squirrel.link/ontology/
[7] https://bioportal.bioontology.org/ontologies/COVIDCRFRAPID
[8] https://github.com/Knowledge-Graph-Hub/kg-covid-19
[9] https://zenodo.org/record/3765375#.XraWJmgzbIU

Knowledge graphs are widely recognized both by industry and academia as the state of the market technology for managing big data using FAIR principles (Blumauer 2020).

Knowledge graphs are based on the following W3C standards:

- International Resource Identifiers (IRI)
- Resource Description Framework (RDF/RDFS)
- Web Ontology Language (OWL)
- Semantic Web Rule Language (SWRL)
- SPARQL Protocol and RDF Query Language (SPARQL)

An IRI looks very much like a URL. The primary difference is that URLs typically point to resources that are meant to be displayed in a browser. IRI's are more general than URLs and can describe resources to a finer level of granularity than an HTML page. An IRI can be any resource such as a class, a property, an individual, etc. (DuCharme, 2011)

RDF is the foundation language for describing IRI data as a graph rather than in relational or other types of formats (W3C 2014).

RDFS is layered on top of RDF and provides basic concepts such as classes, properties, and collections (W3C 2014a).

OWL is layered on top of RDFS and provides the semantics for knowledge graphs. OWL is an implementation of Description Logic which is a decidable subset of First Order Logic (W3C 2012). OWL enables the definition of reasoners which are automated theorem provers. OWL reasoners first ensure that an ontology model is consistent. If the model is not consistent the reasoner will highlight the probable source of the inconsistency. If the model is consistent reasoners can then deduce additional information based on concepts described below such as transitivity, inverses, value restrictions, etc. OWL reasoners originated with the KL-One family of knowledge representation languages and successors to KL-One such as Loom. (MacGregor, 1991).

SWRL is a rule-based language that extends OWL reasoners with additional constructs beyond what can be described with OWL's Description Logic language (W3C 2004).

Finally, SPARQL allows federated queries across heterogeneous sources of data. A SPARQL query defines a graph pattern that is matched against the available data sources and returns the data that matches the pattern (DuCharme, 2011).

# 3 METHDOLOGY

This section provides a description of the CODO ontology design and development methodology.

For designing an ontology, there are several methodologies available in the literature. Some of the state-of-the-art popular approaches are METHONTOLOGY (Fernandez et al., 1997), TOVE (Gruninger and Fox, 1995), DILIGENT (Vrandecic et al., 2005), NeOn (Suárez-Figueroa et al, 2012), UPON (De Nicola et al., 2005), YAMO (Dutta et al., 2015), etc. The design approach of CODO has been primarily influenced by YAMO, a step-by-step approach for building a formally defined large-scale faceted ontology. The YAMO methodology also provides a set of ontology design guiding principles which is quite unique. The steps of the CODO ontology design process are displayed in Figure 1 and described below.

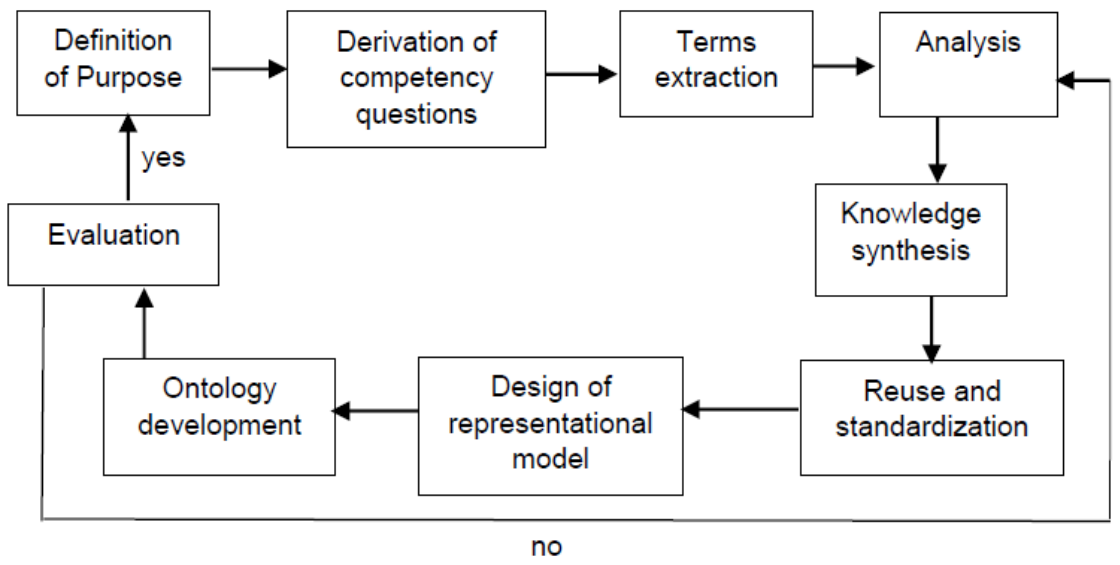

Figure 1: Steps of CODO ontology development process.

**S1**: Definition of purpose - this step describes the purpose and application of the ontology. As discussed above the purpose of the CODO ontology is to facilitate publication of COVID-19 data as a knowledge graph and to develop semantic services and applications (e.g., decision support system, advanced analytics) (Dutta, 2017). Also, to enable various organizations (e.g., government agencies, hospitals, researchers, data publishers, news agencies, etc.) to annotate and describe COVID-19 information.

**S2**: Derivation of competency questions – Elaborate the purpose defined in S1 into a set of competency questions. Some of these competency questions are:

I. How many people recovered from COVId-19 in place p until date t?
II. How many people died in country c?
III. Give me the travel history of patient p.
IV. Give me the COVID-19 patients p and their relationship r, if any.
V. Give me the COVID-19 patients p who are in family relationships f.
VI. Give me the primary reasons i for the maximum number of COVID-19 patients p.
VII. Give me the most prevalent symptoms s of Severe COVID-19 d.
VIII. Find all People p who are related to someone r who has been diagnosed with Covid and who has not yet been tested.

**S3**: Term extraction – in extracting the terms for the ontology, we primarily referred to COVID-19 datasets on cases, patients, relations (e.g., family, co-workers), geographic locations, and date-time information. For this purpose, we referred to data repositories, such as the COVID-19 Data Repository by the Center for Systems Science and Engineering (CSSE) at Johns Hopkins University[10] and the data repository curated by Athreya et al.[11] We also used the literature including government published documents and guidelines. For example, guidance documents on appropriate management of suspected/confirmed cases of COVID-19[12], WHO published literature[13], newspaper articles, etc. on CVID-19. Some of the most significant extracted terms are:

```
patient, doctor, covid-19 dedicated facility, covid care centre, dedicated covid health centre, covid-19 clinical facility, mild and very mild covid-19, severe covid-19, moderate covid-19, exposure to civid-19, vital signs, test finding, symptom, SpO2, cases, blood pressure, temperature
```

**S4**: Analysis - following the extraction of the terms, this steps involves analysing the derived compound and complex concepts and breaking them into their elemental entities. The analysis is done based on the definition and characteristic of each of the concepts and then grouping them according to their similarity. For example, analysing the terms covid care centre (*any facility, such as hotels/lodges/hostels/stadiums for providing care to COVID-19 patients*) and dedicated health centre (*hospitals that shall offer care for all cases that have been clinically assigned as moderate*) based on their definition reveals that both of them have a common point and can be grouped as subclasses of the class for covid dedicated facility.

**S5**: Knowledge synthesis – this step involves synthesizing and arranging the knowledge by defining the relationships between the concepts. This step lead to the discovery of concept hierarchies. For example (the indention indicates the hierarchy)

```
Organization
  COVID-19 dedicated facility
    Covid care centre
    Dedicated covid health centre
    Dedicated covid hospital
```

**S6**: Reuse and standardization – technology can only go so far to enable integration and re-use. Ultimately, what is required is to develop and re-use domain vocabularies. We have followed this best practice in the development of CODO. We have integrated concepts from the following vocabularies into CODO: Schema.org, Friend of a Friend (FOAF) vocabulary[14], SNOMED CT[15] and OBO.[16]

Schema.org is used for modelling common concepts such as gender and locations. FOAF is used to model Agents, such as Person and Organization classes and related properties. SNOMED CT and OBO are used to model clinical findings and symptoms.

**S7**: Design of representational model – involves structuring and modelling the domain knowledge produced in the previous step. The idea is to model the domain knowledge showing its various components, such as classes, properties and their relationships. This is important as in one side it ensures the aggregation, substitution, improvement, sharing and reapplication of the ontology (Dutta et al, 2015, Giunchiglia & Dutta, 2011), and on the other side it provides a consolidated view of the ontology and its coverage. Figure 2 shows a high-level view of the CODO model.

---

[10] https://github.com/CSSEGISandData/COVID-19
[11] https://www.isibang.ac.in/~athreya/incovid19/data.html
[12] https://www.mohfw.gov.in/pdf/FinalGuidanceonMangaementofCovidcasesversion2.pdf
[13] https://www.who.int/news-room/q-a-detail/q-a-coronaviruses
[14] http://xmlns.com/foaf/spec/
[15] http://www.snomed.org/
[16] http://www.obofoundry.org/

**S8**: Ontology development – this step involves developing the formal model using a formal logic language. For developing CODO, we used OWL-DL, a Description Logic ontology language. CODO was designed using the Protégé ontology editor (Musen 2015) developed at Stanford University. In addition to the core editor we utilized the Pellet reasoner, SWRLTab, Cellfie, and Snap SPARQL plugins. The details of the ontology are provided in Section 4.

**S9**: Evaluation – this step involves evaluating how closely the ontology meets the design goals. It gauges the technical competence of the ontology. There is no easy and automatic way of evaluating an ontology. The reasoners can verify the syntactic structure and consistency of the ontology but cannot evaluate the domain knowledge and knowledge structure. The manual evaluation by domain experts is one of the most prevalent methods (Lozano-Tello and Gomez-Perez, 2004, Dutta et al., 2015).
To verify that the CODO ontology serves the purpose it was designed for, we imported data on the pandemic from the government of India using the Cellfie Protégé plugin (described in section 5.1). We also wrote SPARQL queries based on the competency questions described in S2. An example SPARQL query is illustrated in section 5.2.

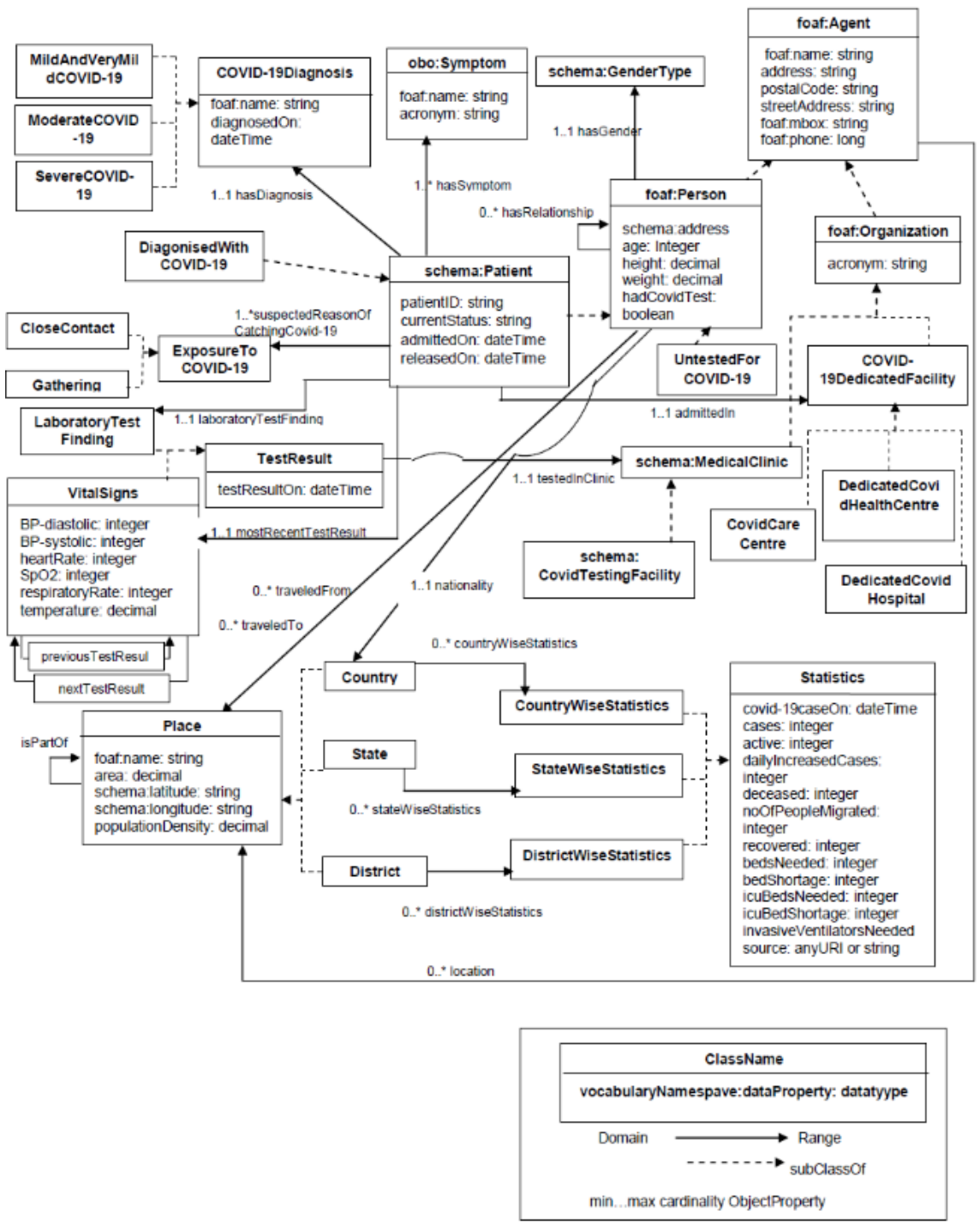

Figure 2: Overview of the CODO model.

# 4 THE CODO ONTOLOGY

In this section we describe some of the important classes, properties, and some sample individuals that we developed to give users of the ontology examples of the types of reasoning that can be automated with the ontology.

The current version CODO1.2 is available here: https://github.com/biswanathdutta/CODO. Also, the HTML specification documents of the ontology is available here: https://isibang.ac.in/ns/codo.

CODO1.2 consists of 50 classes, 62 object properties and 45 data properties. The basic ontology has a handful of sample individuals for illustrative purposes. The first application of CODO on actual data from the government of India has over 23,000 individuals representing cases of the pandemic in India (the data dump is available here: https://github.com/biswanathdutta/CODO).

## 4.1 Properties and Reasoning

One of the main differences between OWL and other object-oriented models is that properties in OWL are first class entities that are not bundled with a specific class. In traditional Object-Oriented Programming (OOP) a property is defined as part of a class definition. If the class is deleted so is the property. In OWL properties are independent entities (W3C 2006).

Properties in OWL are equivalent to binary relations in First Order Logic (FOL). They also have a number of capabilities that relations in FOL have and that can be automatically enforced by an OWL reasoner.

Two examples of such capabilities leveraged by CODO are *symmetric* and *inverse* properties. A symmetric property is such that if the tuple <a, b> is in the property then the tuple <b, a> must be as well. An example of a symmetric property in CODO is hasSpouse. If a Person p000001 hasSpouse p000004 then the reasoner automatically infers that p000004 hasSpouse p000001.

Inverses are defined such that if <a, b> is in a property then <b, a> is in its inverse property. An example of this in CODO are the hasChild and isChildOf properties. These are inverse properties and one merely has to assert that one of the properties holds for two individuals and the reasoner will infer that the appropriate inverse holds for the two individuals as well.

Since OWL properties are FOL relations they are sets (of binary tuples). Thus, just as classes can have subclasses where the subclass is a subset of the superclass so properties can have sub-properties where all the tuples in the super-property are in the sub-property but not necessarily vice versa.

One way this is leveraged in CODO is in the hasRelationship property hierarchy (see figure 3).

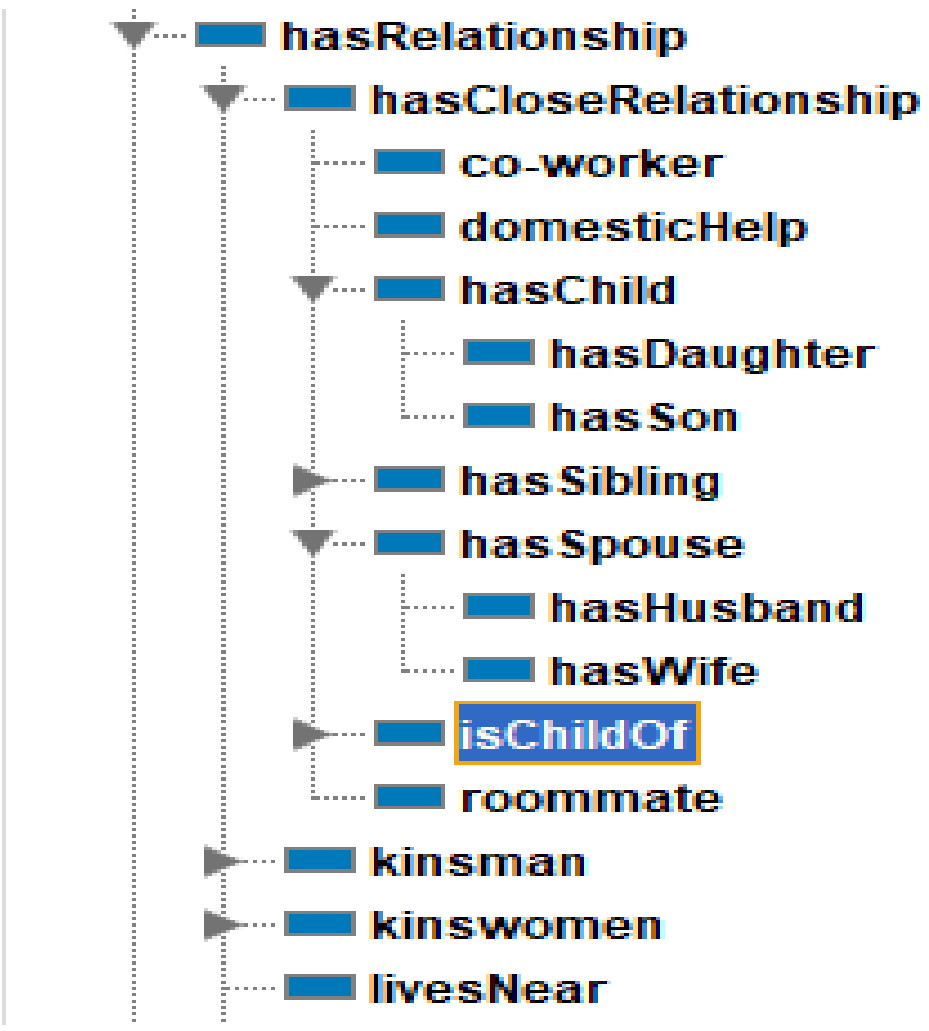

Figure 3: The hasRelationship Property Hierarchy

The hasRelationship property captures some of the ways that people can have interactions with each other. It also has sub-properties that differentiate hasCloseRelationship relations from others. A close relationship is one where the two people are likely to regularly live or work together such as parents and children, co-workers, and roommates. This is distinguished from other types of relationships such as aunts and uncles where it is less likely that the two individuals are in regular close contact.

For example, in the test data for CODO we asserted that p000001 hasDaughter p000007. The reasoner automatically inferred that p000001 hasChild p000007 (because hasDaughter is a sub-property of hasChild) and that p000001 hasCloseRelationship p000007 (because hasChild is a sub-property of hasCloseRelationship). This property hierarchy will be leveraged further as we combine it with the capability to define necessary and sufficient axioms for classes in the next section.

## 4.2 Defined Classes

OWL can be used to define axioms that are necessary and sufficient for an individual to be a member of a class. The OWL reasoners can use these axioms to automatically restructure the class hierarchy as well as to do significant additional reasoning about individuals.
If one defines axioms for a class in the SubClassOf field in Protégé these are necessary axioms for the class. I.e., they must be true for any individual that is a member of that class but it may not be the case that every individual that fulfils that axiom is a member of that class. When axioms are defined in the EquivalentTo field in Protégé these axioms are both necessary and sufficient conditions for that class. I.e., any individual that satisfies those axioms is automatically inferred to be an instance of that class. Classes with necessary and sufficient axioms are known as defined classes in OWL. In CODO we have combined sub-properties with a defined class to create a defined subclass of Person called UrgentlyNeedsCovidTest. The *necessary and sufficient axioms* for this class are:

```
foaf:Person
 and (hasCloseRelationship some
DiagnosedWithCovid) and (hadCovidTest value
false)
```

DiagnosedWithCovid is also a defined class with necessary and sufficient conditions such that anyone who has been diagnosed with the virus is a member of that class. Thus, UrgentlyNeedsCovidTest defines a class for anyone who has a close relationship that has been diagnosed with Covid-19 and who has themselves not yet had a Covid-19 test.

Figure 4 displays this defined class. The individuals in the instances field are instances of the Person class that the reasoner has inferred are also instances of this defined class. Note: anything in Protégé highlighted in yellow was not defined by some input data but was inferred by the reasoner based on the data and the axioms in the ontology.

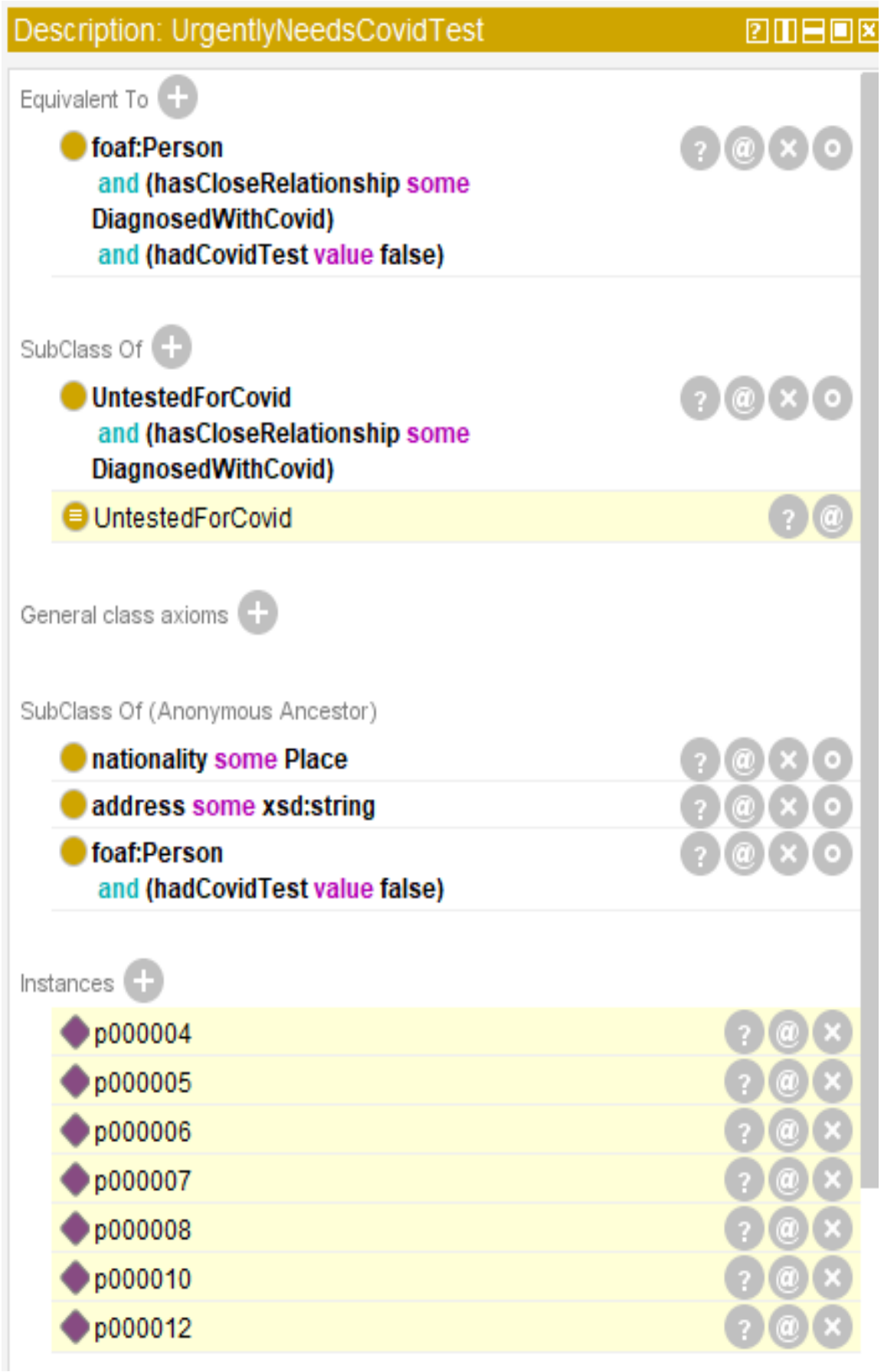

Figure 4 UrgentlyNeedsCovidTest Defined Class.

## 5. CODO Evaluation

In this section we describe how we have evaluated the CODO ontology. This evaluation was done by:

- Populating the ontology with real world data from the government of India.
- Making the ontology available as a vocabulary for others to use which has already occurred with one system on the Bioportal site.
- Developing SPARQL queries which implement some of the use cases identified in our development methodology.
- Exporting the ontology to a commercial triplestore product which provides capabilities for much larger datasets and additional visualization.

### 5.1 Data population

In this section we describe how we have populated the ontology with data from the Indian Ministry of Health and Family Welfare website.[17] This data has been collected into spreadsheets by Siva Athreya and other researchers at the Indian Statistical Institute.[18] A snapshot of a datasheet is shown in Figure 5.

To integrate this data into the ontology we used the Cellfie Protégé plugin (O'Connor 2010). Cellfie allows the user to define transformation rules to convert rows in a spreadsheet into instances of a class in an ontology and property values for that instance (see Table 1 for an example of a transformation rule).

We utilized Cellfie to import data about the pandemic from the Indian province of Karnataka. Each row in the spreadsheet corresponded to a case where a specific patient was diagnosed with Covid. In the CODO ontology each row was transformed into an individual of the Patient class and values in each row such as the age, sex, date of diagnosis, etc. were transformed into the appropriate property values for each patient.

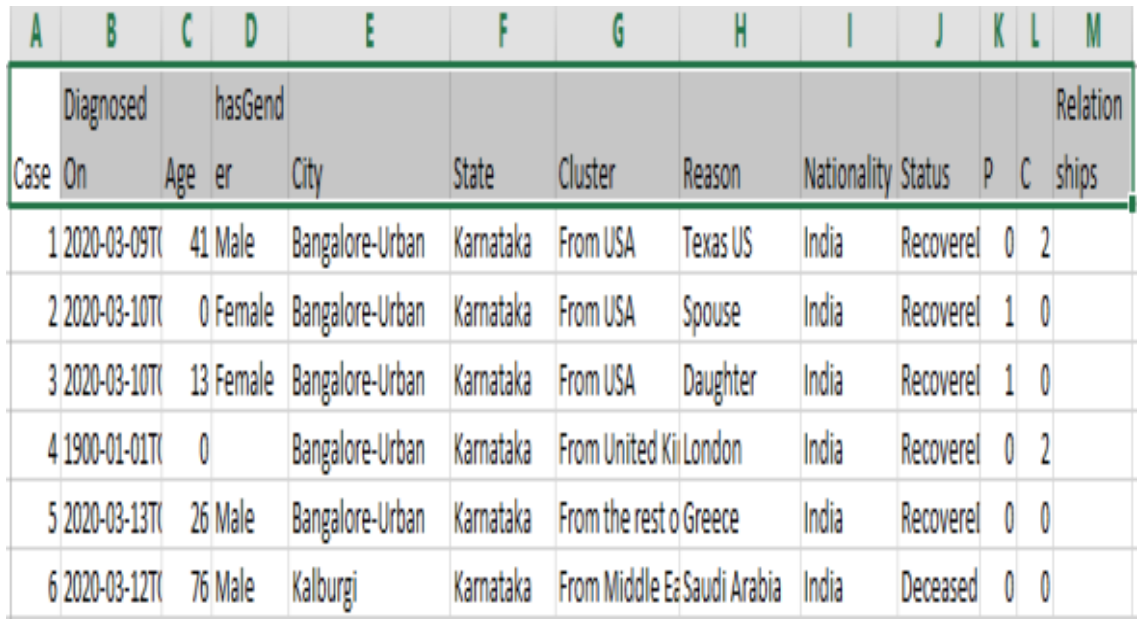

| A | B | C | D | E | F | G | H | I | J | K | L | M |
|---|---|---|---|---|---|---|---|---|---|---|---|---|
| Case | Diagnosed On | Age | hasGender | City | State | Cluster | Reason | Nationality | Status | P | C | Relationships |
| 1 | 2020-03-09T( | 41 | Male | Bangalore-Urban | Karnataka | From USA | Texas US | India | Recoverel | 0 | 2 | |
| 2 | 2020-03-10T( | 0 | Female | Bangalore-Urban | Karnataka | From USA | Spouse | India | Recoverel | 1 | 0 | |
| 3 | 2020-03-10T( | 13 | Female | Bangalore-Urban | Karnataka | From USA | Daughter | India | Recoverel | 1 | 0 | |
| 4 | 1900-01-01T( | 0 | | Bangalore-Urban | Karnataka | From United Ki | London | India | Recoverel | 0 | 2 | |
| 5 | 2020-03-13T( | 26 | Male | Bangalore-Urban | Karnataka | From the rest o | Greece | India | Recoverel | 0 | 0 | |
| 6 | 2020-03-12T( | 76 | Male | Kalburgi | Karnataka | From Middle Ea | Saudi Arabia | India | Deceased | 0 | 0 | |

Figure 5: A glimpse of the dataset.

The resulting ontology had over 23,000 individual patients with data from March to the beginning of July 2020.

Table 1: Example Transformation rule.

```
Individual: @A*(mm:hashEncode
rdfs:label=("patient", @A*))
   Types: Patient
   Facts: 'diagnosed on' @B*(xsd:dateTime),
              age @C*(xsd:decimal),
              'has gender' @D*,
               'city' @E*,
              'state' @F*,
               'travelled from' @G*,
               nationality @I*,
```

[17] https://www.mohfw.gov.in/

[18] https://www.isibang.ac.in/~athreya/incovid19/data.html

```
            status @J*,
            'has caused any secondary infections'
@L*(xsd:boolean)
```

## 5.2 SPARQL Queries

The SPARQL query engine is roughly analogous to OWL as SQL is to relational databases. However, since the underlying structure of OWL are graphs rather than tables, SPARQL constructs graph patterns and then searches knowledge graphs for any individuals that match the graph pattern. Like SQL, SPARQL can do more than query, it can also delete, insert, and transform data (DuCharme, 2011).

SPARQL has many features that provide additional value beyond the capabilities described so far. For one thing, SPARQL can integrate data from multiple heterogeneous data sources. The beginning of each SPARQL query starts with a list of namespaces and the IRI where these namespaces can be found. Hence, SPARQL can do queries across broad data sets from multiple sources enabling a truly federated virtual knowledge base. Since different data sources may have different formats SPARQL can use pattern matching to transform data from various sources.

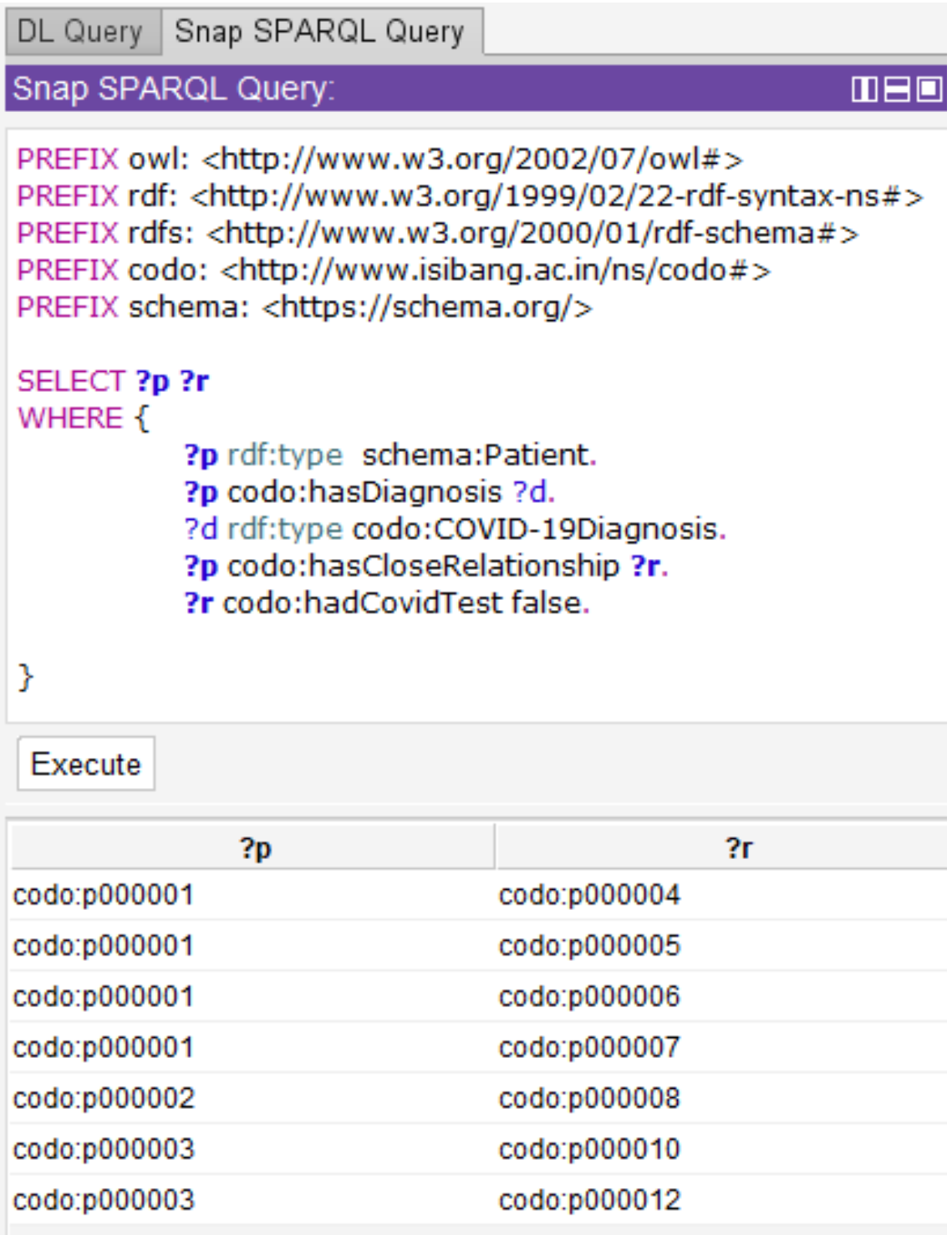

Figure 6: CODO SPARQL Query.

Figure 6 displays a SPARQL query using the Snap SPARQL query plugin in Protégé. The above shows the SPARQL syntax for the query "*Find all People who have a close relation to someone who has been diagnosed with Covid and who has not yet been tested.*"

The Prefixes first define the various namespaces that the query will utilize and their IRIs. In this case the query performs the same logic as the defined class described in section 4.2. One advantage of using the SPAQRL query is that in addition to seeing the specific individuals who match the query (the ?r column) we can also see the closely related individual that has been diagnosed with Covid-19 (the ?p column).

## 5.3 Utilization of CODO Vocabulary

One of the primary design goals for CODO was that it could serve as a reusable vocabulary for other projects. Although we have only recently published the ontology on Github and Bioportal, we already have one user from the Bioportal site: the Ping COVID-19 risk detection system.[19]

## 5.4 Triplestores and Visualization

Protégé is a modelling tool not a persistent storage tool. Although it is possible to persist knowledge graphs designed in Protégé with small to medium sets of test data, to achieve the true power of knowledge graph technology a triplestore product is required. A triplestore is a database designed to store data as graphs rather than as relational tables (Blumauer 2020).

Although the current test data in CODO can be stored in files from Protégé we have already begun to hit the limits of Protégé with the data we have imported from the Indian government. We have begun to utilize a triplestore environment in anticipation of scaling up CODO to having data for up to a million patients rather than the thousands currently in the ontology. We have imported CODO into the free version of the Allegrograph triplestore product from Franz Inc. The free version is still capable of supporting 5 million triples and also supports Allegro's Gruff visualization tool. Figure 7 displays a small number of test data patients from the current CODO ontology using Allegro's Gruff tool.

[19] http://bioportal.bioontology.org/projects/Ping

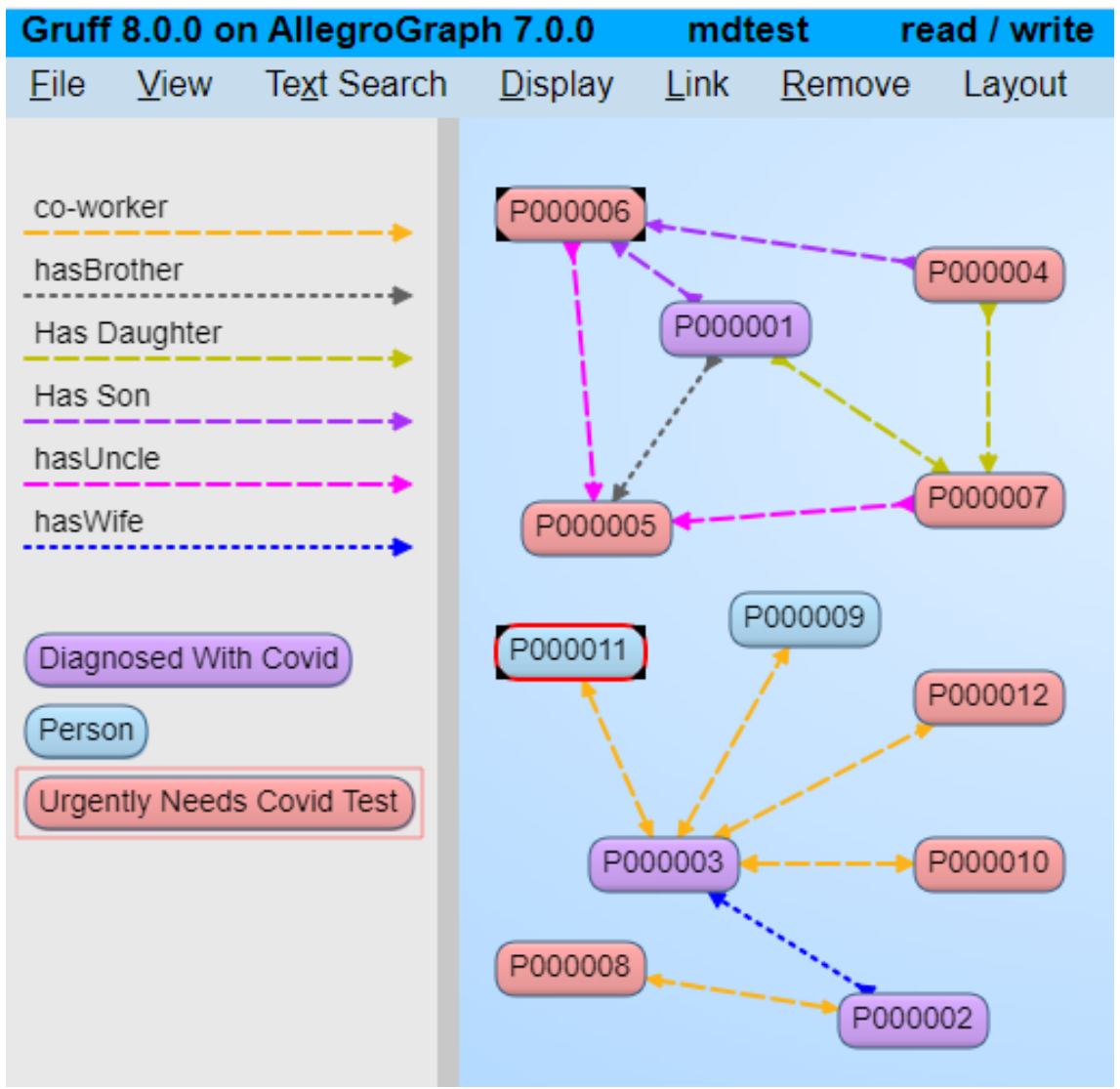

Figure 7: CODO in Allegro's Gruff Visualization Tool.

## 6 CONCLUSIONS

The CODO ontology is only the first step in providing a knowledge graph model for COVID-19 based on FAIR data principles. The current CODO ontology has already found its use in a real world project called Ping and in uploading thousands of cases from data collected by the government of India. The main limitation of the current work is it lacks a truly rigorous evaluation of the developed ontology. In our future work, we aim to evaluate the ontology by health domain experts and also by applying the Information Retrieval system evaluation technique. In addition, we plan to enhance the current CODO ontology by integrating many more COVID-19 datasets available on the Web, both from India and world-wide. Finally, we plan to publish CODO using a triplestore database published as a SPARQL endpoint. This will provide capabilities to handle much larger datasets. It will also enable SPARQL queries that can integrate CODO with other complimentary ontologies such as CIDO.

## ACKNOWLEDGEMENTS

This work was conducted using the Protégé resource, which is supported by grant GM10331601 from the National Institute of General Medical Sciences of the United States National Institutes of Health. Thanks to Franz Inc. (http://www.allegrograph.com) and its help with AllegroGraph and Gruff.